\documentclass[12pt,reqno]{amsart}
\allowdisplaybreaks[4]
\usepackage{graphicx} 
\usepackage{amssymb}
\usepackage{amsmath}
\usepackage{cite}

\numberwithin{equation}{section}
\makeatletter      
\@addtoreset{equation}{section}
\makeatother       
\newtheorem{proposition}{Proposition}
\newtheorem{corollary}{Corollary}
\newtheorem{lemma}{Lemma}

\begin{document}

\title[Additional Symmetries and String Equation]{
NEGATIVE GENERATORS OF THE VIRASORO CONSTRAINTS FOR THE BKP
HIERARCHY}
\author{Jingsong He\dag\ddag, Kelei Tian\dag, Angela Foerster\ddag }
\dedicatory { \dag\ Department of Mathematics, USTC, Hefei, 230026 Anhui, P.\ R.\ China\\
\ddag\ Instituto de F\'{\i}sica da UFRGS, Av. Bento Gon\c{c}alves
9500, Porto Alegre, RS - Brazil }

\begin{abstract}

We give a straightforward derivation of the string equation and
Virasoro constraints on the $\tau$ function of the BKP hierarchy by
means of some special additional symmetry flows. The explicit forms
of the actions of these additional symmetry flows on the wave
function and then the negative Virasoro generators $L_{-k}$ are
given, where $k$ is a positive integer.

\end{abstract}

 \maketitle



PACS(2003):\ 02.30.Ik


\section{Introduction}
Since its introduction in a very convenient form in 1986\cite{os1}  and
stimulated  by the importance of the string equation\cite{douglas1}, much
attention has been paid to the study of the additional symmetries
and the Virasoro constraints\cite{morozov1,asv1,dl2} for the Kadomtsev-Petviashvili(KP)
hierarchy\cite{dkjm0,dl1}. For this hierarchy, it is now well established that there
are two representations of the additional symmetries, i.e. Sato
vertex operator form \cite{dkjm0} and Orlov -Schulmann(OS) M-operator form\cite{os1}.
In 1994, they were proved to be equivalent  by the
action on the wave functions of the KP hierarchy in the two
different forms\cite{asv2,asv3}. In this process, Adler-Shiota-van
Moerbeke(ASvM) formula plays a crucial role. Almost at the same time
Dickey presented a very elegant and compact proof of ASvM formula
\cite{dl3} based on the Lax operator $L$ and OS's M operator. In addition
he also \cite{dl4} derived the string equation, the action of the additional
symmetries on the $\tau$ function and the Virasoro constraints of
the KP hierarchy.

The BKP hierarchy\cite{dkjm0,dkjm3} is  a reductional sub-hierarchy of the KP with a restriction
on the Lax operator $L^*=- \partial L \partial^{-1} $ (here $*$ stands for a formal adjoint
operation, $L$ is a Lax operator of the BKP hierarchy). Therefore, it was natural to expect
intensive investigations  on the additional symmetry and its associated structures
of the BKP hierarchy after the discovery of this kind of symmetries on the KP hierarchy.
In this context, Johan\cite{vdl1,vdl2} has obtained  the Virasoro constraints on the $\tau$
function and ASvM formula of the BKP hierarchy by an algebraic method. Takasaki\cite{kt1} found
appropriate restrictions on the generators of the additional symmetries for the BKP hierarchy.
Very recently, by using Takasaki's result \cite{kt1} and Dickey's method \cite{dl4},
Tu\cite{tu1} has given an
explicit form of the generators of the additional symmetries, and then an alternative proof
of the ASvM formula for the BKP hierarchy. This new proof  is more simpler and
transparent in comparison with the algebraic method presented in \cite{vdl2}. Here it is
important
to mention that due to the reductional conditions of the BKP hierarchy many differences in
relation to the KP hierarchy may emerge, turning this investigation highly non-trivial. For
example,  the generators of the additional symmetries of the BKP hierarchy  are also
correspondingly restricted\cite{kt1}(specifically,see eq.(40)). This fact implies that the generators
of the additional symmetries for the  BKP  hierarchy  must be different compared to their
counterparts on the KP hierarchy. In this scenario, it would be relevant  to derive the Virasoro
constraints of the BKP hierarchy using also the potentialities of the Dickey's method.

In this work,  applications of the additional symmetries of the BKP
hierarchy are studied in detail. In particular, we find the string
equation and negative generators of Virasoro constraints on the
$\tau$ function for the BKP hierarchy by means of the additional
symmetry flows. We also give the explicit forms of the negative
Virasoro generators by calculating the action of the additional
symmetry flows on the $\tau$ function, which is induced by the
action of the additional symmetry flows on the wave function of the
BKP hierarchy.

The organization of this paper is as follows. In section 2  we
present a brief summary of the BKP and its additional symmetry,
which is followed by string equation and some special additional
symmetry flow equations in section 3. In section 4 we derive the
negative generators of the Virasoro constraints. Section 5 is
devoted to conclusions and discussions.

\section{BKP Hierarchy and its additional symmetries}
Let $L$ be the pseudo-differential operator,
\begin{equation}\label{KPlaxoperator}
L=\partial +u_1\partial^{-1}+u_2\partial^{-2}+
u_3\partial^{-3}+\cdots,
\end{equation}
and then the KP hierarchy is  defined by the set of partial differential equations $u_i$
with respect to independent variables $t_j$
\begin{equation}\label{KPhierarchy}
\dfrac{\partial L}{\partial {t_n}}=[B_n, L],\ n=1,2,3, \cdots.
\end{equation}
Here $B_n=(L^n)_+=\sum\limits_{k=0}^n a_k\partial^k$  denotes the
non-negative powers of $\partial $ in $L^n$, $\partial
=\partial/\partial_x$, $u_i=u_i(x=t_1,t_2,t_3,\cdots,)$. The other
notation $L^n_{-}=L^n-L^n_+$ will be needed by the sequent text. $L$
is called the Lax operator and eq.(\ref{KPhierarchy}) is called the
Lax equation of the KP hierarchy. In order to define the BKP
hierarchy, we need a formal adjoint operation $*$ for an arbitrary
pseudo-differential operator $P=\sum\limits_i p_i \partial^i$,
$P^*=\sum\limits_i (-1)^i\partial^i  p_i$. For example,
$\partial^*=-\partial$, $(\partial^{-1})^*=-\partial^{-1}$, and
$(AB)^*=B^*A^*$ for two operators. The BKP hierarchy\cite{dkjm0,dkjm3} is a
reduction of the KP hierarchy by the constraint
\begin{equation}\label{BKPconstraint}
 L^*=- \partial L \partial^{-1} ,
\end{equation}
which compresses all even flows of the KP hierarchy, i.e. the Lax
equation of the BKP hierarchy has only odd flows ,
\begin{equation}\label{BKPhierarchy}
\dfrac{\partial L}{\partial {t_{2n+1}}}=[B_{2n+1}, L],\ n=0,1,2, \cdots.
\end{equation}
Thus $u_i=u_i(t_1,t_3, t_5,\cdots)$ for the BKP hierarchy.

  The Lax equation of the BKP hierarchy can be given by the consistent conditions of the
following set of linear partial differential equations
\begin{equation}\label{CKPlinearsystem}
Lw(t,\lambda)=\lambda w(t,\lambda),\dfrac{\partial w(t,\lambda)}{\partial {t_{2n+1}}}
=B_{2n+1}w(t,\lambda), \ t=(t_1,t_3,t_5, \cdots).
\end{equation}
Here $w(t,\lambda)$ is identified as a wave function. Let $\phi$ be
the wave operator(or Sato operator) of the BKP hierarchy
$\phi=1+\sum_{i=1}^{\infty}w_i\partial^{-i}$, then the Lax operator
and the wave function admit the following representation
\begin{equation}\label{satorepresenation}
L=\phi \partial \phi^{-1}, \ \ w(t,\lambda)=\phi(t)e^{\xi(t,\lambda)}=\hat{w}e^{\xi(t,\lambda)},
\end{equation}
in which $\xi(t,\lambda)=\lambda t_1
+\lambda^3t_3+\cdots+\lambda^{2n+1}t_{2n+1}+\cdots$,
$\hat{w}=1+\frac{w_1}{\lambda}+\frac{w_2}{\lambda^2}+\frac{w_3}{\lambda^3}+\cdots$.
Similar to the KP hierarchy, the BKP hierarchy also has a sole
function,$\tau$ function $\tau (t)=\tau (t_1,t_3,t_5,
\cdots,t_{2n-1} \cdots) $ ($n$ is a positive integer), such that all
of the dynamical coordinates $u_{i}$ can be expressed, and further the
wave function is

\begin{equation}\label{taurepresenation}
w(t,\lambda)=\hat{w}(t,\lambda)e^{\xi(t,\lambda)}=\frac{\tau (t_1
-\frac{2}{\lambda},t_3-\frac{2}{3\lambda ^3},t_5-\frac{2}{5\lambda
^5}, \cdots)}{\tau (t)}e^{\xi(t,\lambda)} \equiv \frac{\tilde{\tau}
(t,\lambda)}{\tau(t)} e^{\xi(t,\lambda)}
\end{equation}
It is easy to show that the Lax equation is equivalent to Sato equation
\begin{equation}\label{satoequation}
\dfrac{\partial \phi}{\partial{t_{2n+1}}}= -L^{2n+1}_{-} \phi,
\end{equation}
and the constraint on $L$ in eq.(\ref{BKPconstraint}) is transformed to the
constraint on the wave operator
\begin{equation}\label{BKPconstraintwaveoperator}
\phi^*=\partial \phi^{-1} \partial^{-1}.
\end{equation}
Eq.(\ref{BKPconstraintwaveoperator}) is a crucial condition to
construct the additional symmetries of the BKP hierarchy, which will
affect the action of the additional symmetry on  the operator $\phi$. It leads to
a distinct explicit form of the generators of the additional symmetry
in comparison to the cases of the KP hierarchy \cite{os1,dl1} and the CKP hierarchy
\cite{htma1}, as we shall see latter.

Now we recall the additional symmetries given by Tu\cite{tu1} of the BKP
hierarchy. Let the OS's operator $M$ be given by
\begin{equation}\label{Moperator}
M=\phi\Gamma \phi^{-1},\
\Gamma=\sum\limits_{i=1}^{\infty}(2i-1)t_{2i-1}\partial^{2i-2}=t_1+3t_3\partial^2
+5t_5\partial^4+\cdots,
\end{equation}
then they satisfy the useful technical identities
\begin{equation}\label{MLL}
[M,L^l]=-lL^{l-1},l\in Z,
\end{equation}
\begin{equation}\label{MML}
[M^m,L]=-mM^{m-1},m\in Z_+.
\end{equation}
Define the additional flows
\begin{equation}\label{definitionadditionalflowsonphi}
\dfrac{\partial \phi}{\partial
{t^*_{m,l}}}=-\left(A_{m,l}\right)_{-}\phi,
\end{equation}
or equivalently
\begin{equation}\label{definitionadditionalflowsonLandM}
\dfrac{\partial L}{\partial{t^*_{m,l}}}=-[(A_{m,l})_{-}, L],
\end{equation}
where $A_{m,l}=A_{m,l}(L,M)$ are monomials in $L$ and $M$. As pointed
in the last paragraph, constraints on $L$ in eq.(\ref{BKPconstraint}), or equivalently
on $\phi$ in eq.(\ref{BKPconstraintwaveoperator}) imply
restrictions on the generators, and then one distinct form of $A_{m,l}$ \cite{tu1} is
\begin{equation}\label{AML}
A_{m,l}=M^mL^l-(-1)^lL^{l-1}M^mL.
\end{equation}
Indeed, this generator is different compared to  results
$A_{m,l}=M^mL^l$\cite{os1,dl1} for the KP hierarchy and $
A_{m,l}=M^mL^l-(-1)^lL^{l}M^m$\cite{htma1} for the CKP hierarchy.
\begin{proposition}
([16])1) The additional flows are symmetries of the BKP hierarchy.
2)They form a centerless $W_{1+\infty}^B$-algebra understanding
their actions on $\phi$ as eq.(\ref{definitionadditionalflowsonphi}).
\end{proposition}
\section{Some special additional symmetry flow equations}
We further concentrate on some special additional symmetry flows in
order to find suitable additional flows implying the Virasoro
constraints on the $\tau$ function of the BKP hierarchy. So two
examples are calculated in the following.
\begin{proposition}
The action on $L$ of the additional flows associated
with $A_{1,l} =-(l-1)L^{l-1}$ is in the form of
\begin{equation}
\partial_{t^*_{1,l}}L=(l-1)[(L^{l-1})_-,L]=\left\{\begin{array}{l}0,\ for\ l=0,-2,-4,-6,\cdots.\\
-(l-1)(\partial_{t_{1-l}}L),\  for\ l =
2,4,6,\cdots.\end{array}\right.
\end{equation}
\end{proposition}
Although this result is different with its counterpart in the KP
hierarchy, this case is not interesting enough because this
additional symmetry flows are almost equivalent to the CKP flows
acing on the space of the Lax operators $L$. The reason is that $l$
is an even integer.

Therefore we consider  $A_{1,-(l-1)}$, and calculate its action on
$L^l$. For this end, from now on assume that $l = 2k$ and $k$ is a
positive integer. By using eq.(\ref{MLL}), the $A_{1,-(l-1)}$ can be
expressed as
\begin{equation}
A_{1,-(l-1)} = 2ML^{-(l-1)}-lL^{-l},
\end{equation}
and then
\begin{equation}\label{eqaddsymm1lminus1onLl}
\begin{array}{l}\hspace{2cm}\partial_{t^*_{1,-(l-1)}}L^l
=-[(A_{1,-(l-1)})_-, L^l]\\ = [(A_{1,-(l-1)})_+, L^l] +
[-(A_{1,-(l-1)}), L^l] = [(A_{1,-(l-1)})_+, L^l] + 2l.\end{array}
\end{equation}
Thus we get the following proposition based on the actions of the
additional symmetry $A_{1,-(l-1)}$ on the $L^l$.
\begin{proposition}\label{propaddsymmonLl}
 Let $l=2m(2n+1)$,  $m,n=1,2,3,\cdots$,  and $L^l$ is independent of
$t^*_{1,-(l-1)}$, then the string equation of the BKP hierarchy is
\begin{equation}\label{eqconstraintonLl}
[L^l, \frac{1}{2l}(A_{1,-(l-1)})_+] = 1.
\end{equation}
  Furthermore, this equation can be written in
a more explicit form as follows,
\begin{equation}\label{eqformofconstraintonLl}
[L^{2k},\frac{1}{ 2k} ML^{-(2k-1)}-\frac{1 }{2} L^{-2k}] = 1, \
k=m(2n+1).
\end{equation}
\end{proposition}

{\bf Proof} The eq.(\ref{eqaddsymm1lminus1onLl}) and $\partial_{t^*_{1,-(l-1)}}L^l = 0$
deduce directly eq.(\ref{eqconstraintonLl}). Moreover $\partial_{t^*_{1,-(l-1)}}L^l =
0$ infers $(A_{1,-(l-1)})_- = 0$, and then $(ML^{-(2k-1)})_- =
kL^{-2k}$ and $(A_{1,-(2k-1)})_+ = 2ML^{-(2k-1)}-2kL^{-2k}$. Taking
this back into eq.(\ref{eqformofconstraintonLl}), then eq.(\ref{eqformofconstraintonLl})
is proved, which completes the proof.\hfill $\square$

 Note that eq.(\ref{eqformofconstraintonLl})
was also obtained by Johan\cite{vdl1} from the Virasoro constraints
on the $\tau$ function of the BKP hierarchy. However, his equation
is not the string equation without the restrictions of $l$. In other
words, $L^l$ can not equal $(L^l)_+$ with an arbitrary positive even
integer $l$.
\begin{corollary}
\label{coraddflowonL2k}
If $L^l$  satisfy the eq.(\ref{eqconstraintonLl}),then
\begin{equation}\label{eqaddflowonL2k}
-\frac{1}{2k}\sum_{n\geq k+1} (2n- 1)t_{2n-1}(\partial
_{t_{2n-(2k+1)}}L^{2k}) = 1.
\end{equation}
\end{corollary}
Let k = 1, the zero order terms of above equation tell us
\begin{equation}\label{eqaddflowonL2kwithkisone}
\frac{1}{2}\sum_{n\geq2} (2n-
1)t_{2n-1}(\partial _{t_{2n-3}}\tau) + \frac{1}{8}x^2\tau = 0.
\end{equation}

This result is indeed distinct with the case of KP hierarchy given
by corollary 1.2 of Ref. \cite{av1}.

\noindent{\bf Proof} By a direct calculation, the left hand side of
eq.(\ref{eqconstraintonLl}) becomes $$1 =\left [L^{2k}, \frac{1}{2k}
(ML^{-(2k-1)})_+\right ]= \left [L^{2k}, \frac{1}{2k}\left
(\phi\sum_{n=1}^{\infty}(2n-1)t_{2n-1}\partial^{2n-2k-1}\phi^{-1}\right)_+\right
]$$$$=\left [L^{2k}, \frac{1}{2k}\left (\phi\sum_{n\geq
k+1}^{\infty}(2n-1)t_{2n-1}\partial^{2n-2k-1}\phi^{-1}\right
)_+\right ].\hspace{4cm}$$
 Note that the change in index of
summation is due to the identity $(\phi\partial^{-m}\phi^{-1})_+ =
0,$ here $m$ is a positive integer. We also should note $L^k_+ =
(\phi\partial^k\phi^{-1})_+$ with $k\geq0$, and then get $$1 = \left
[L^{2k}, \frac{1}{2k}\sum_{n\geq
k+1}^{\infty}(2n-1)t_{2n-1}(L^{2n-2k-1})_+\right
]=-\frac{1}{2k}\sum_{n\geq
k+1}^{\infty}(2n-1)t_{2n-1}(\partial_{t_{2n-2k-1}}L^{2k}),$$ which
is eq.(\ref{eqaddflowonL2k}). Furthermore, let $k = 1$, taking $L^{2k} = L^2 =
\partial^2 +2u_1+$lower order terms and $u_1 = 2(\ln\tau)_{xx}$ back into eq.
(\ref{eqaddflowonL2k}), we get the zero order terms in both sides,
$$-\frac{1}{2}{\sum_{n\geq2}} (2n - 1)t_{2n-1}(4\partial_{ t_{2n-3}} (\ln\tau)_{xx}) = 1.$$
By exchanging the order of the derivative with respect to $x$ and $t_{2n-3}$, then
 $$-2
{\sum_{n\geq2}} (2n - 1)t_{2n-1}\left(\frac{1}{\tau}(\partial_{
t_{2n-3}}\tau)\right)_{xx} = 1.\hspace{0.5cm}$$
Integrating the above
formula two times on $x$ and choosing suitable constants, then
eq.(\ref{eqaddflowonL2kwithkisone}) is reached, and thus completes the proof.\hfill $\square$
\section{Virasoro generators}

It is easy to find that eq.(\ref{eqconstraintonLl}) is equivalent to $\partial_{t^*_{
1,-(l-1)}}\phi= 0,$ with $l = 2k$. To get the Virasoro constraints on
the $\tau$ function and the Virasoro generators, firstly we shall pass the
action of the flows $\partial_{t^*_{ 1,-(l-1)}}$ on the wave operator
$\phi$ to the action on the wave function $w$, and then on the
$\tau$ function of the BKP hierarchy. In this context, $\hat{w}(t,
z)$ plays the role of a bridge connecting actions on the wave operator
$\phi$ and on the $\tau$ function. The following lemmas are
necessary to do this.

 \begin{lemma}
\label{lemA1lminus1}
 For $l=2k,\ k = 1, 2, 3, 4,
\cdots$,
$$(A_{1,-(l-1)})_- = 2\phi\left(\sum_{n=1}^{k}(2n-1)t_{2n-1}\partial^{2(n-k)-1}\right )
\phi^{-1}$$
\begin{equation}
\hspace{2cm}+2\sum_{n=k+1}^{\infty}(2n-1)t_{2n-1}L^{2(n-k)-1}_{-}-lL^{-l}
\end{equation}
\end{lemma}

{\bf Proof} According to the definitions of $M$ and $L$,
$$(ML^{-(l-1)})_- =(\phi\Gamma\phi^{-1}\phi\partial^{-(l-1)}\phi^{-1})_{-}=(
\phi\Gamma\partial^{-(l-1)}\phi^{-1})_-\hspace{7cm}$$
$$\hspace{0.7cm}=
\left (\phi \big(\sum_{n=1}^{k}(2n-1)t_{2n-1}\partial^{2n-2k-1}
+\sum_{n=k+1}^{\infty}(2n-1)t_{2n-1}\partial^{2n-2k-1}
\big)\phi^{-1}\right )_-$$
$$\hspace{1.8cm}= \phi\left
(\sum_{n=1}^{k}(2n-1)t_{2n-1}\partial^{2n-2k-1}\right
)\phi^{-1}+\left
(\sum_{n=k+1}^{\infty}(2n-1)t_{2n-1}\phi\partial^{2n-2k-1}\phi^{-1}\right
)_-$$
$$\hspace{0.3cm}= \phi\left
(\sum_{n=1}^{k}(2n-1)t_{2n-1}\partial^{2n-2k-1}\right
)\phi^{-1}+\sum_{n=k+1}^{\infty}(2n-1)t_{2n-1}L^{2(n-k)-1}_-$$ In
the second term of the last second equality, $\phi$ pass the
$t_{2n-1}$ because $\phi$ is involved only with $\partial_{t_1}$,
but there $2n - 1 > 1$. Thus, taking this representation of
$(ML^{-(l-1)})_-$ into the generator $A_{1,-(l-1)} = 2ML^{-(l-1)} -
lL^{-1}$, and then the lemma is proved.\hfill $\square$
\begin{proposition}\label{propaddfolwsonwhat}
Let $l = 2k$ as
before, and $\hat{w}(t, z)$ is given by eq.(2.7), then
$$\partial_{t^*_{ 1,-(l-1)}} \hat{w}(t,z) = -2\left (z^{-2k+1}(
\frac{\partial}{\partial z}\hat{w}) + \sum_{n=1}^{k} (2n -
1)t_{2n-1}z^{2n-2k-1}\hat{w}\right )$$
\begin{equation}\label{eqaddfolwsonwhat}
\hspace{1cm}+ 2\sum_{n=k+1}^{\infty} (2n -
1)t_{2n-1}\frac{\partial\hat{w}}{\partial t_{2n-2k-1}} + 2kz^{-2k}
\hat{w}.
\end{equation}
\end{proposition}
 {\bf Proof} First of all, by using lemma \ref{lemA1lminus1},
the additional symmetry flow acts on the wave operator $\phi$ as
$$\partial_{t^*_{1,-(l-1)}} \phi= -(A_{1,-(l-1)})_-\phi= -2\phi \sum_{n=1}^{k}(2n -
1)t_{2n-1}\partial^{2(n-k)-1}\hspace{2cm}$$
$$ \hspace{4.5cm}+
2\sum_{n=k+1}^{\infty} (2n -
1)t_{2n-1}(\partial_{t_{2(n-k)-1}}\phi) + l\phi\partial^{-l}.$$
Note that this is an operator equation, thus we can apply the function $e^{xz}$ to both side
simultaneously. Therefore, by
applying to both sides of the last formula $e^{xz}$ and using two
identities: $[\phi, x]e^{xz} = (\frac{\partial}{\partial
z}\hat{w})e^{xz}$ and $\phi\partial^{-l}e^{xz}=\phi z^{-l}e^{xz} =
z^{-l}\hat{w}e^{xz}$, we achieve that
$$(\partial_{t^*_{1,-(l-1)}}
\hat{w})e^{xz}=
-((A_{1,-(l-1)})_-\phi)e^{xz}\hspace{7cm}$$
$$\hspace{0.7cm}=
-2\phi(xz^{-2k+1}e^{xz}+ \sum_{n=2}^{k}(2n -
1)t_{2n-1}z^{2n-2k-1}e^{xz})$$$$ \hspace{1.7cm}+
2\sum_{n=k+1}^{\infty} (2n -
1)t_{2n-1}(\partial_{t_{2(n-k)-1}}\hat{w})e^{xz}
+2kz^{-2k}\hat{w}e^{xz}$$ $$\hspace{1.2cm}=
-2(z^{-2k+1}(\frac{\partial}{\partial z}\hat{w})+ \sum_{n=2}^{k}(2n
- 1)t_{2n-1}z^{2n-2k-1}\hat{w})e^{xz}$$$$ \hspace{1.7cm}+
2\sum_{n=k+1}^{\infty} (2n -
1)t_{2n-1}(\partial_{t_{2(n-k)-1}}\hat{w})e^{xz}
+2kz^{-2k}\hat{w}e^{xz}.$$
Dividing  from above equality the factor $e^{xz}$, the
result of the proposition is obtained, and thus completes the
proof.\hfill $\square$

Further, we know from proposition \ref{propaddfolwsonwhat} that the equivalent form of
eq.(\ref{eqconstraintonLl}), $\partial_{t^*_{1,-(l-1)}}\phi= 0$, implies the
constraints on wave function, $\partial_{t^*_{1,-(l-1)}}\hat{w}= 0$,
specifically. So it is very natural to express these constraints on
the wave function by means of the $\tau$ function. By proceeding in  this way, the
explicit form of the Virasoro generators will be obtained as
follows.
\begin{lemma}\label{lemshiftfun}
The action of additional symmetries
on $\hat{w}$ can be expressed as a special form of
\begin{equation}\label{eqshiftfun}
(\partial_{t^*_{ m,n}}\hat{w}) = f(t;z)\frac{\tilde{
\tau}(t,z)}{\tau(t)},
\end{equation}
where $f(t; z) =
g_1(t_{2n-1}\rightarrow t_{2n-1 }- \frac{2}{(2n-1)z^{2n-1}}
;\tilde{\tau}(t, z)) - g_1(t;\tau(t)) \equiv \tilde{g}(t; z) -
g(t)$, which is called a similarity shifted function.
\end{lemma}
\noindent{\bf Proof} By a straightforward calculation, we
have$$(\partial_{t^*_{ m,n}}\hat{w}) = \left (\partial_{t^*_{
m,n}}\frac{\tilde{ \tau}(t,z)}{\tau(t)}\right
)=\frac{\tau(t)\partial_{t^*_{
m,n}}\tilde{\tau}(t,z)-\tilde{\tau}(t,z)\partial_{t^*_{
m,n}}\tau(t)}{\tau^2(t)}$$$$\hspace{1.6cm}=\left
(\frac{\partial_{t^*_{
m,n}}\tilde{\tau}(t,z)}{\tilde{\tau}(t,z)}-\frac{\partial_{t^*_{
m,n}}\tau(t)}{\tau(t)}\right )\frac{\tilde{
\tau}(t,z)}{\tau(t)}=f(t;z)\frac{\tilde{ \tau}(t,z)}{\tau(t)},
$$ as required. This is the end of the proof.\hfill $\square$

This lemma reminds us that we should transform the $(\partial_{t^*_{1,-(l-1)}}\hat{w})$
given by eq.(\ref{eqaddfolwsonwhat}) to be the form of eq.(\ref{eqshiftfun}) in
order to find $(\partial_{t^*_{1,-(l-1)}}\tau(t))$, and then find
Virasoro generators. However, in general, only this is not enough to
guarantee us to get the correct Virasoro generators. Normally, the form
of $(\partial_{t^*_{ m,n }}\hat{w})$ is not unique, as pointed out by
Dickey \cite{dl4} for the KP hierarchy, which is also true for the BKP
hierarchy, because there exists a freedom of gauge transformation with
constant coefficients. One has to choose a suitable gauge such that
a good form of $(\partial_{t^*_{ m,n }}\hat{w})$ can be reached, and
then the latter can lead to the correct Virasoro generators by means
of $(\partial_{t^*_{ m,n }}\tau)$. In particular, for the case of $n
\geq0$, it is very complicated to get a general simple expression of
$(\partial_{t^*_{1,n+1}}\hat{w})$ for the KP hierarchy\cite{dl4} and BKP hierarchy.
However, it is simpler for the $n < 0$. This is a reason for us to
only study $(\partial_{t^*_{ 1,-(l-1)}}\hat{w})$ in the last
proposition \ref{propaddfolwsonwhat} and to study $(\partial_{t^*_{ 1,-(l-1)}}\tau(t))$ in the
sequent proposition with $l\geq2$.


\begin{lemma}\label{Virasorogenerator}Suppose
\begin{equation}\label{eqvirasorogeneratorsgivenbylemma}
L_{-k}=\frac{1}{2} \sum_{n=k+1}^{\infty} (2n - 1)t_{2n-1}
\frac{\partial}{
\partial t_{2n-2k-1}} + \frac{1}{8}\sum_{n+m=k+1}(2n - 1)(2m -
1)t_{2n-1}t_{2m-1},
\end{equation}
where $n$ and $m$ in the second term take value from 1 to
$k$ under the condition of $n+m = k+1$, then the Virasoro commutation relations
\begin{equation}
[L_{-k},L_{-l}]=(-k+l)L_{-(k+l)}
\end{equation}
hold  for integers $k,l\geq 1$.
\end{lemma}


\begin{proposition}\label{propvirasorogenerator} If $L^l$ satisfies eq.(3.4),
i.e., $L$ is independent of $t^*_{ 1,-(l-1)}$, $l = 2k,\ k = 1, 2,
3,\cdots$, then the Virasoro constraints imposed on the $\tau$
function of the BKP hierarchy are
\begin{equation}\label{eqvirasorogenerator}
L_{-k}\tau = 0,
\end{equation}
where $L_{-k}$ are Virasoro
generators given by eq.(
\ref{eqvirasorogeneratorsgivenbylemma}).
\end{proposition}
Obviously, let $k = 1,\ L_{-1}\tau= 0$ is consistent with the
corollary \ref{coraddflowonL2k}. The $\dfrac{1}{k}L_{-k}$  gives the same result as that obtained
in Ref.\cite{vdl1} using a completely different approach. In particularly, the Virasoro generators for
the BKP are indeed different with ones \cite{fkn} of the  KP hierarchy.

\noindent{\bf Proof} For convenience, we mark the four terms of
$(\partial_{t^*_{ 1,-(l-1)}}\hat{w})$ in eq.(\ref{eqaddfolwsonwhat}) by $a),\ b),\ c),\
d)$, respectively. The proof has the following steps. In this
proof, $\tilde{\tau} =\tilde{\tau} (t, z)$.

\noindent1). First of all, we try to construct the similarity
shifted function structure in two terms, $a)$ and $c)$, because only
they have the derivatives of $\tau$ and $\tilde{\tau}$. A direct
calculation shows
\begin{equation}
a)\equiv-2z^{-2k+1}\frac{\partial\hat{w}}{\partial z}=
 -2z^{-2k+1}\frac{1}{\tau}\frac{\partial\tilde{\tau}}{\partial z}
 = - \frac{4}{\tau}\sum_{ n=k+1}^{\infty}\frac{1}{z^{2n-1}}
 \frac{\partial\tilde{\tau}}{\partial t_{2n-2k-1}} ,
\end{equation}

and
\begin{align*}
c) & \equiv 2
\sum_{n=k+1}(2n-1)t_{_{2n-1}}(\dfrac{\partial}{\partial
t_{_{2n-2k-1}}}\dfrac{\tilde{\tau}}{\tau}) \\
& =
\dfrac{2}{\tau}\sum_{n=k+1}^{\infty}(2n-1)t_{_{2n-1}}\dfrac{\partial
\tilde{\tau}}{\partial t_{_{2n-2k-1}}}-\dfrac{2
\tilde{\tau}}{\tau^2}\sum_{n=k+1}^{\infty}(2n-1)t_{_{2n-1}}\dfrac{\partial
\tau}{\partial t_{_{2n-2k-1}}}.
\end{align*}
For the third term, we try to make a similarity shifted function
deliberately by insertion of one term, and thus
\begin{align}
c)&
=(\dfrac{2}{\tilde{\tau}}\sum_{n=k+1}^{\infty}(2n-1)(t_{_{2n-1}}-\dfrac{2}{(2n-1)z^{2n-1}})\dfrac{\partial
\tilde{\tau}}{\partial
t_{_{2n-2k-1}}}-\dfrac{2}{\tau}\sum_{n=k+1}^{\infty}(2n-1)t_{_{2n-1}}\dfrac{\partial
\tau}{\partial t_{_{2n-2k-1}}})\dfrac{\tilde{\tau}}{\tau}   \notag \\
& +
\dfrac{4}{\tau}\sum_{n=k+1}^{\infty}\dfrac{1}{z^{2n-1}}\dfrac{\partial
\tilde{\tau}}{\partial t_{_{2n-2k-1}}}.
\end{align}
Further,
\begin{align}\label{eqac}
a)+c)= &
\Big(\dfrac{2}{\tilde{\tau}}\sum_{n=k+1}^{\infty}(2n-1)(t_{_{2n-1}}-\dfrac{2}{(2n-1)z^{2n-1}})\dfrac{\partial
\tilde{\tau}}{\partial t_{_{2n-2k-1}}} \notag \\
&
-\dfrac{2}{\tau}\sum_{n=k+1}^{\infty}(2n-1)t_{_{2n-1}}\dfrac{\partial
\tau}{\partial t_{_{2n-2k-1}}}\Big)\dfrac{\tilde{\tau}}{\tau}
\end{align}
possess one similarity shifted function as we expected.  \\
2)For the b)+d), it can not be transformed to a similarity shifted
function if we only make a shift $t_{_{2n-1}}\rightarrow
t_{_{2n-1}}-\frac{2}{(2n-1)z^{2n-1}}$ in b), similar to what like we have done in
c). So we have to try it by using its product, as the second
simplest case. To do this, by rewriting b) as a symmetrical form,
and then using one identity, we get
\begin{align}
b)\equiv &-2
\sum_{n=1}^{k}z^{2n-2k-1}(2n-1)t_{_{2n-1}}\hat{w}   \notag \\
= &-
(\mbox{\hspace{-0.3cm}}\sum_{n+m=k+1}\mbox{\hspace{-0.4cm}}(2n-1)(2m-1)t_{_{2n-1}}\dfrac{\hat{w}}{(2m-1)z^{2m-1}}
+\mbox{\hspace{-0.4cm}}\sum_{n+m=k+1}\mbox{\hspace{-0.4cm}}(2n-1)(2m-1)t_{_{2m-1}}\dfrac{\hat{w}}{(2n-1)z^{2n-1}})    \notag \\
= &
-\dfrac{1}{2}\Big(-\sum_{n+m=k+1}(2n-1)(2m-1)(t_{_{2m-1}}-\dfrac{2}{(2m-1)z^{2m-1}})(t_{_{2n-1}}-\dfrac{2}{(2n-1)z^{2n-1}}) \notag \\
& + \sum_{n+m=k+1}(2m-1)(2n-1)t_{_{2n-1}}t_{_{2m-1}}\Big)\hat{w} -
2 \sum_{n+m=k+1}\dfrac{\hat{w}}{z^{2(n+m)-2}}   \notag \\
= &
\dfrac{1}{2}\Big(\sum_{n+m=k+1}(2m-1)(2n-1)(t_{_{2m-1}}-\dfrac{2}{(2m-1)z^{2m-1}})(t_{_{2n-1}}-\dfrac{2}{(2n-1)z^{2n-1}})  \notag \\
& -
\sum_{n+m=k+1}(2m-1)(2n-1)t_{_{2n-1}}t_{_{2m-1}} \Big)\hat{w}-2kz^{-2k}\hat{w}.
\end{align}
Therefore,
\begin{align}\label{eqbd}
b) + d) = & \dfrac{1}{2}\Big(\mbox{\hspace{-0.3cm}}\sum_{n+m=k+1}\mbox{\hspace{-0.4cm}}(2n-1)(2m-1)(t_{_{2m-1}}-\dfrac{2}{(2m-1)z^{2m-1}})(t_{_{2n-1}}-\dfrac{2}{(2n-1)z^{2n-1}}) \notag \\
& -
\sum_{n+m=k+1}(2m-1)(2n-1)t_{_{2n-1}}t_{_{2m-1}}\Big)\dfrac{\tilde{\tau}}{\tau}.
\end{align}
3) Taking a) + c) in eq.(\ref{eqac}) and b) + d) in eq.(\ref{eqbd}) into
$(\partial_{_{t_{_{1,-(l-1)}}^*}}\hat{w})$ in eq.(\ref{eqaddfolwsonwhat}), then
\begin{align}\label{eqabcd}
(\partial_{_{t_{_{1,-(l-1)}}^*}}\hat{w}) = &
\Big(\dfrac{2}{\tilde{\tau}}\sum_{n=k+1}^{\infty}(2n-1)(t_{_{2n-1}}-\dfrac{2}{(2n-1)z^{2n-1}})
\dfrac{\partial
\tilde{\tau}}{\partial t_{_{2n-2k-1}}} \notag \\
& \mbox{\hspace{-2.4cm}}+
\dfrac{1}{2}\sum_{n+m=k+1}\mbox{\hspace{-0.4cm}}(2n-1)(2m-1)(t_{_{2n-1}}-\dfrac{2}{(2n-1)z^{2n-1}})(t_{_{2m-1}}-
\dfrac{2}{(2m-1)z^{2m-1}})\Big)
\dfrac{\tilde{\tau}}{\tau} \notag \\
& \mbox{\hspace{-2.4cm}}-
\Big(\dfrac{2}{\tau}\sum_{n=k+1}^{\infty}(2n-1)t_{_{2n-1}}\dfrac{\partial
\tau}{\partial t_{_{2n-2k-1}}} +
\dfrac{1}{2}\sum_{n+m=k+1}(2n-1)(2m-1)t_{_{2n-1}}t_{_{2m-1}}\Big)
\dfrac{\tilde{\tau}}{\tau}
\end{align}
On the one side, taking into account an equivalent form of the
eq.(\ref{eqconstraintonLl}), i.e., $\partial_{_{t_{_{1,-l-1}}^*}}\phi=0$, and the lemma
\ref{lemshiftfun}, we have
\begin{equation}\label{eqthesecondformaddflowsonwhat}
\partial_{_{t_{_{1,-(l-1)}}^*}}\hat{w}  = \Big(\dfrac{\partial_{_{t_{_{1,-l-1}}^*}} \tilde{\tau}}{\tilde{\tau}}-
\dfrac{\partial_{_{t_{_{1,-l-1}}^*}} \tau}{\tau}\Big)
\dfrac{\tilde{\tau}}{\tau}=0,
\end{equation}
and then deduce
\begin{equation}\label{eqaddflowsontau}
(\partial_{_{t_{_{1,-l-1}}^*}} \tau)=c_{_{1}}\tau.
\end{equation}
where $c_{_{1}}$ is a constant. On the other side, comparing
eq.(\ref{eqabcd}) and eq.(\ref{eqthesecondformaddflowsonwhat}) infers
\begin{align}\label{eqthesecondformaddflowsontau}
(\partial_{_{t_{_{1,-(l-1)}}^*}} \tau) = &
\Big(2\hspace{-0.3cm}\sum_{n=k+1}^{\infty}(2n-1)t_{_{2n-1}}\dfrac{\partial
\tau}{\partial
t_{_{2n-2k-1}}}+\dfrac{1}{2}\sum_{n+m=k+1}\hspace{-0.3cm}(2n-1)(2m-1)t_{_{2n-1}}t_{_{2m-1}}\tau\Big)+c_{_{2}}\tau  \notag \\
= & 4L_{_{-k}}\tau + c_{_{2}} \tau
\end{align}
with an arbitrary constant $c_{_{2}}$, and $L_{_{-k}}$ as we
expected. Therefore, the Virasoro constraints on the $\tau$ function
\begin{equation*}
L_{_{-k}} \tau = 0
\end{equation*}
is obtained from eqs.(\ref{eqaddflowsontau}) and (\ref{eqthesecondformaddflowsontau}) with $c_{_{1}}=c_{_{2}}$.
This is the end of the proof. \hfill $\square$

\section{Conclusions and Discussions}
We have studied the applications of the additional symmetry flows of the
BKP hierarchy, and thus provided the following main results:
\begin{itemize}
\item 1) the action of the special additional symmetry flows on $L^{l}$ and the string equation
in proposition \ref{propaddsymmonLl};
\item 2) the explicit forms of the actions of the additional symmetry
flows on the wave function  $(\partial_{{t_{_{1,-(l-1)}}^*}}\hat{w})$ in proposition \ref{propaddfolwsonwhat};
\item 3) the explicit forms of the negative Virasoro generators and the Virasoro
constraints on the $\tau$ function of the BKP hierarchy in proposition \ref{propvirasorogenerator}.
\end{itemize}
In addition, the similarity shifted function $f(t;z)$ in
$(\partial_{{t_{_{m,n}}^*}}\hat{w})$ given by lemma \ref{lemshiftfun} is also
crucial to  get the action of the additional symmetry flows
on the $\tau$ function. Our route
is \textit{additional symmetry} $\rightarrow$  \textit{additional
symmetry flow equations eq.(\ref{eqconstraintonLl}) associated with $A_{1,-(l-1)}$}
$\rightarrow$\textit{actions of the additional symmetry flows on the wave function }
$\rightarrow$ \textit{Virasoro constraints on the $\tau$ function}
$\rightarrow$ \textit{Virasoro generators}.

\par
For the further research related to this topic, the extension of the
additional symmetry and its associated structures for the
multi-component BKP hierarchy \cite{kt2} would be  very  interesting
and relevant although very  complicated. Moreover, it is also an
interesting problem to calculate out the whole set of  $L_{k}$ for
the Virasoro constraints and $W_n$ for the $W$-constraints for the
BKP hierarchy.


{\bf Acknowledgments} {\small This work is supported by the NSF of
China under Grant No.10301030 and No.10671187, and SRFDP of China
under Grant No.20040358001. Support of the joint post-doc fellowship
of TWAS(Italy) and CNPq(Brazil)at UFRGS is gratefully acknowledged.
J.He thanks Professors LiYishen, ChengYi (USTC, China) and F.
Calogero(University of Rome ``La Sapienza",Italy) for long-term
encouragements and supports. J.He also thanks Professor K.
Takasaki(Kyoto University,Japan) for his kind clarifying some
questions on his new paper by Email.}




\end{document}